\documentclass[twocolumn, preprintnumbers, amsmath, amssymb, prb, showpacs, floatfix]{revtex4}
\usepackage[dvips]{graphicx}
\usepackage{bm}
\usepackage{shapepar}

\begin{document}

\title{Spin injection, accumulation and spin precession in a mesoscopic non-magnetic metal island}

\author{M. Zaffalon}
\author{B.J. van Wees}
\affiliation{Department of Applied Physics and Materials Science Centre\\
University of Groningen, Nijenborgh 4, 9747 AG Groningen, the Netherlands}

\begin{abstract}
We experimentally study spin accumulation in an aluminium island with all dimensions smaller than the spin relaxation length, so that the spin imbalance throughout the island is uniform. Electrical injection and detection of the spin accumulation are carried out in a four terminal geometry by means of four cobalt electrodes connected to the island through tunnel barriers. We model the system theoretically and we investigate the role of the ferromagnetic electrodes on the spin accumulation at the limit at which the electron diffusion time can be neglected. We present measurements of spin accumulation at room temperature and at 4.2~K: in both cases the spin accumulation signal is larger than the Ohmic resistance of the aluminium island. From magnetisation precession measurements at room temperature, we extract a spin relaxation time $\tau_{sf} = 60$~ps and a polarisation $P = 8\%$ for tunnel barriers with resistances as low as 20~$\Omega\mu\mathrm{m}^2$. We show that the precession measurements are invariant under the interchange of voltage and current electrodes, and under the reversal of magnetic fields and magnetisations, according to the reciprocity theorem. We show that spin accumulation and spin precession in a system with uniform magnetisation can be described in terms of the (relative) orientation of the ferromagnetic contacts' magnetisations and we determine from precession measurements the angles between the magnetisation direction of the contacts.
\end{abstract}

\maketitle

\section{INTRODUCTION}
Creating and manipulating a non-equilibrium magnetisation in a non-magnetic metal is a central requirement in the field of spintronics\cite{awschalom}. The orientation of an electron spin injected in a non-magnetic metal is the result of the interaction of the spin intrinsic magnetic moment with the magnetic fields in which the spin is moving. In the presence of a uniform magnetic field, the spin precesses coherently around the field's direction and its orientation changes with a uniform precession frequency.

In a diffusive metal, a (non-uniform) effective magnetic field arises from the relativistic motion of the spin in the electric field of the metal ions and the defects\cite{elliot}. This is called spin orbit interaction and it is responsible for the randomisation of the spin orientation: the relaxation of the non-equilibrium magnetisation occurs by transferring the spin angular momentum to the metal lattice in a time scale in the order of 100~ps.

To induce a spin current and a spin accumulation, a spin dependent scattering is required for the conduction electrons. The usual approach in all-electrical transport experiments is to drive a current from a ferromagnet (FM) whose band structure is spin dependent, to a non-magnetic (N) metal. It has been noted that the main obstacle for efficient spin injection in the diffusive regime is the short spin relaxation length in the ferromagnet, a problem known by the name of conductance mismatch\cite{rashba}. In GMR experiments\cite{GMR}, with vertical devices, the useful signal can be made large enough for practical applications by reducing the distance between the ferromagnetic layers. In lateral structures, this is not a feasible solution, due to technological limitations. Also, in the clean contact regime, measures have to be taken to ensure that either the current path is perpendicular to the FM/N interface or that the voltage probes are separate from the current path. Anomalous magneto-resistance and Hall effect can mimic and hide the spin accumulation\cite{jedema_01}.

In order to overcome these problems, tunnel barriers were proposed\cite{rashba} at the interface between the ferromagnetic metal and the normal metal layer (the tunnel barrier conductance being proportional to the FM density of states), thereby making the tunnel barrier the dominant (spin dependent) resistance of the system.

Previous experiments have studied the spin current in systems larger than the spin relaxation length. A seminal experiment was performed by Johnson and Silsbee\cite{johnson_85} in the clean contact regime and four-terminal configuration, in a device with two lateral dimensions larger than the spin relaxation length. The experiment was performed on single crystal aluminium bar. The long spin relaxation length they found, $\lambda_{sf} = 50~\mu$m at 4.2~K, allowed them to observe a weak (tens of pV) spin precession signal at macroscopic scale. In diffusive metallic systems, with typical relaxation lengths in the $\mu$m range, observations were done by Jedema \textit{et al.}\cite{jedema_01, jedema_02} at room temperature in a one-dimensional device, both with clean contacts and with tunnel barriers at the FM/N interfaces in four-terminal devices. The spin signal in the clean case was about 1~m$\Omega$ and about two orders of magnitude larger for the devices with tunnel barriers, proving their efficiency as spin injector/detector. Spin accumulation occurs in two terminal pillar structures with all dimensions shorter than the spin relaxation length, used to study the magnetisation reversal of a thin FM layer, driven by a spin polarised current created by a second massive FM layer\cite{katine}. The torque exerted on the FM layer is proportional to the spin accumulation, which is in the mV range.

Recently\cite{zaffalon} we have performed electrical injection and detection of spin accumulation in an aluminium island with all lateral dimensions shorter than the spin relaxation length $\lambda_{sf} = \sqrt{D\tau_{sf}}$ ($\approx 0.6~\mu$m at RT), where $D$ is the diffusion constant and $\tau_{sf}$ the spin relaxation time. Tunnel barriers separated the island from the cobalt electrodes. As opposed to the previous four-terminal experiments, the spins in our case are confined to the island, and since the diffusion time $\tau_{dif\!f} = L^2 / D$ ($L$ being the island's size) is shorter than $\tau_{sf}$, the induced magnetisation behaves uniformly within the island, so that the spatial variation of the magnetisation can be disregarded and the system is zero-dimensional with respect to the spin.

The description of coherent spin transport in a diffusive metal is obtained from the Boltzmann transport equation. The two-channel (spin up and spin down) model of Valet and Fert\cite{VF}, which successfully describes the experimental results of giant magneto-resistance, is however limited to the collinear (magnetisation either parallel or anti-parallel) case. In the general case, one has to retain all the information about the spin direction inside the bulk metal and at the FM/N interfaces.

A theoretical approach to systematically studying the transport through FM/N interfaces in the non-collinear situation was developed by Brataas \textit{et al.}\cite{brataas}. The relevant parameter, alongside the interface conductance for spin up and spin down electrons $G^\uparrow, G^\downarrow$, is a (complex valued) mixing conductance term $G^{\uparrow \downarrow}$, describing the reflection of electron spins perpendicular to the magnetisation of the ferromagnet. $G^{\uparrow\downarrow}$ is related  to the amount of angular moment that the electron spin has transferred to the ferromagnet, and plays an important role in the description of the spin torque and spin pumping\cite{tserkovnyak, heinrich}. 

Here we present a systematic study of non-collinear spin accumulation in a small metallic island, extending the results of Ref.~\onlinecite{zaffalon}. Theoretically, we apply the circuit theory to the system and show how the presence of the FM contacts provides an additional (and anisotropic) mechanism for the relaxation of the spin accumulation, with the relaxation occurring at a faster rate in the direction perpendicular to the FM magnetisation axis.

In Sec.~\ref{theory}, we derive a formula in the zero dimensional limit ($\tau_{dif\!f} \ll \tau_{sf}$) for the spin accumulation as a function of the contacts properties. Sec.~\ref{sample_fabrication} describes the sample fabrication and Sec.~\ref{this_experiment} is a short summary of the relevant theory. The geometry of the sample and the measuring configurations are described in Sec.~\ref{meas_configuration}.

Sec.~\ref{results} presents an extensive set of measurements of spin accumulation in a non-magnetic island at 4.2~K and at room temperature (RT). We use spin precession as a tool to analyse the spin accumulation and extract the relevant parameters such as the spin relaxation time, and the direction of the magnetisation of the FM contacts. We also show that, for our device, the magnetisation relaxation is independent of the mixing conductance term.

\section{THEORY}
\label{theory}

To completely characterise the electronic transport in the linear regime, including the spin, four chemical potentials are necessary, one spin independent (charge) and three spin dependent (magnetisation). The main idea of the finite element theory of Brataas \textit{et al.}\cite{brataas}, is to divide the system into (normal or ferromagnetic) \textit{nodes} connected to each other or to reservoirs by \textit{contacts} (interfaces). A contact can be spin selective, that is it can have different conductances for the two spin populations.

In each node, spin accumulation appears as a result of the spin currents through the contacts. In turn, the amount of spin current through each contact is determined by its conductance and by the chemical potentials of the two nodes on each side of the contact.

Thus, the problem of the transport in the system is broken down into the solution of the motion of charge and magnetisation inside a node with the additional boundary conditions given by the charge and spin current through the contacts. The finite element theory of Brataas \textit{et al.} provides an elegant way to describe the charge and spin currents through FM/N interfaces.

\begin{figure}

\begin{picture}(0, 0)
\put(0, -80){\LARGE $\mu, \bm{\mu}$}
\end{picture}

\includegraphics[width = 0.8\linewidth]{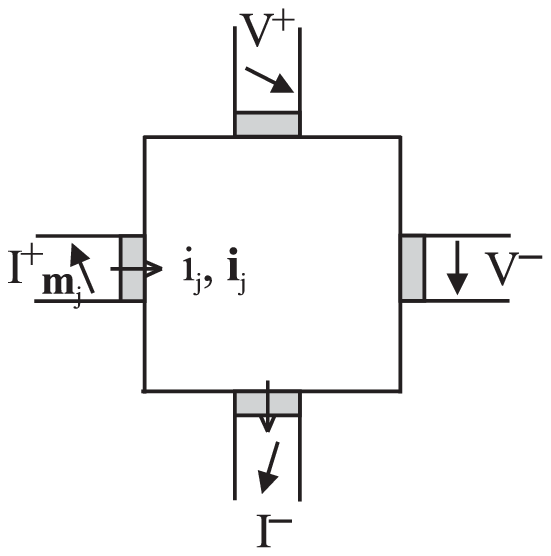}
\includegraphics[scale = 1]{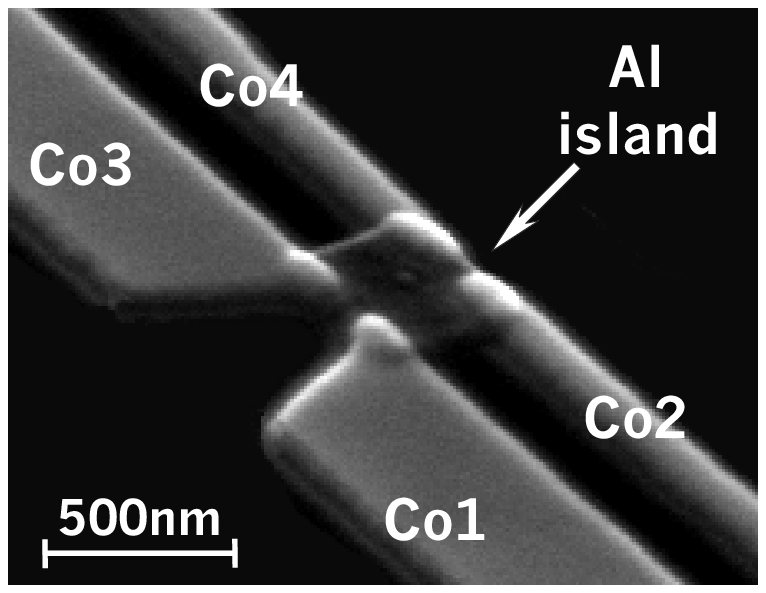}
\caption{(top) A schematic representation of the system: the node (the square island of non-magnetic metal) is characterised by the spin independent and dependent chemical potentials $\mu,\ \bm{\mu}$, connected to ferromagnetic electrodes. $\mathbf{m}_j$ is the unit vector parallel to the magnetisation of the FM. The shadowed regions represent the contacts, separating the node from the reservoirs. (bottom) Scanning electron microscope micrograph of the real device. The square aluminium island is connected to four cobalt electrodes through transparent tunnel barriers\cite{sem}.}
\label{island}
\end{figure}

In the following, we will briefly review the elements of the theory that are relevant for our experimental situation. We will then present an analytical solution for the situation in which the electron diffusion time inside the island $\tau_{dif\!f}$ can be neglected when compared to the spin relaxation time $\tau_{sf}$. This last assumption is equivalent to saying that the distribution of magnetisation is uniform in the island and we will therefore call it, a zero-dimensional system from here on.

In a FM/N device, and for arbitrary configuration of the magnetic reservoirs, a spin current injected from a FM reservoir to a node of normal metal will cause a non equilibrium accumulation of magnetisation. In the linear regime, we describe the transport properties by means of four chemical potentials $\mu(x), \bm{\mu}(x)$, where $\mu(x) = \int f(\epsilon) \mathrm{d}\epsilon$ and $\bm{\mu} = \int \mathbf{f}(\epsilon) \mathrm{d}\epsilon$ are the spin independent and spin dependent chemical potentials in the node and $f(\epsilon)$ and $\mathbf{f}(\epsilon)$ are the spin independent and spin dependent distribution functions. $\bm{\mu} = (\mu_x, \mu_y, \mu_z)$ represents the spin accumulation in different directions and $|\bm{\mu}|$ its magnitude. In equilibrium, no spin accumulation exists in normal and ferromagnetic nodes or reservoirs.

Only in the particular case in which the spin accumulation has the same direction throughout the entire system (and the contacts are all collinear), one can use a description in terms of two, spin up and spin down, chemical potentials: these are related to $\mu, \bm{\mu}$ by $\mu_\uparrow = \mu + |\bm{\mu}|$ and $\mu_\downarrow = \mu - |\bm{\mu}|$.

In a reservoir, the spin independent chemical potential is set by the applied bias voltage $eV$.

In the experimental device, see Fig.~\ref{island}, the node is an aluminium island 400~nm$\times$400~nm$\times$30~nm, the four cobalt electrodes act as reservoirs and the Al$_2$O$_3$ tunnel barriers at the aluminium/cobalt interface are the contacts. The motion of charge and magnetisation in the island is diffusive as the mean free path is of the order of 5~nm at RT and 20~nm at 4.2~K.

The finite element theory specifies the charge and spin \textit{particle} currents $i$ and $\mathbf{i}$ through the contacts that connect the island. This is related to the amplitude probabilities $r^{nm}_\uparrow$ for the reflection of a spin up electron from mode $m$ to mode $n$ in the normal metal, evaluated at the normal side of the contact. If, as in our case, the tunnel barrier is non-magnetic and $\mathbf{m}_j$ is the unit vector representing the magnetisation direction of the $j$ electrode, the natural choice for the quantisation axis is collinear to $\mathbf{m}_j$ (and $\uparrow$ means, for instance, parallel to $\mathbf{m}_j$ and $\downarrow$ anti-parallel). $G^\uparrow_j$ and $G^\downarrow_j$ are the conductances for the up and down spin channels, $G_j = G_j^\uparrow + G_j^\downarrow$ is the total contacts' conductance and $P_j = (G^\uparrow_j - G^\downarrow_j) / (G^\uparrow_j + G^\downarrow_j)$ the polarisation of the interface.

Also the assumption that spin transport through the contacts can be specified only in terms of $r^{nm}_\uparrow$, $r^{nm}_\uparrow$ implies that a spin up electron have zero probability of being converted into a spin down electron, that is no spin flips occur inside the contacts\cite{brataas}.

It is also assumed that spin accumulation in the FM side can only be collinear to the magnetisation direction, i.e. the spin dependent chemical potential is of the form $\bm{\mu}^F = |\bm{\mu}^F| \mathbf{m}$, as the large exchange field rapidly randomises the spin component perpendicular to $\mathbf{m}$.

The charge current entering the normal metal reads\cite{FM_acc}

\begin{equation}
e^2 i_j = G_j\left (\mu^F_j - \mu\right ) - P_jG_j \left (\mathbf{m}_j \cdot \bm{\mu} - |\bm{\mu}^F_j|\right ) \\
\label{eq:charge_current}
\end{equation}

and the spin current

\begin{align}
e^2 \mathbf{i}_j =&
P_j G_j(\mu^F_j - \mu) \mathbf{m}_j + G_j \left (\bm{\mu}^F_j - (\mathbf{m}_j\cdot \bm{\mu}) \mathbf{m}_j\right ) \nonumber \\
+& 2ReG^{\uparrow \downarrow}_j \mathbf{m}_j \times (\mathbf{m}_j \times \bm{\mu}) - 2ImG^{\uparrow \downarrow}_j \mathbf{m}_j \times \bm{\mu}
\label{eq:spin_current}
\end{align}

$\mu^F_j$ being the spin independent chemical potential of the FM electrode $j$. The conductances are defined according to the Landauer-B\"uttiker formalism:

\begin{equation}
G^\uparrow = \frac{e^2}{h}\left [M - \sum_{nm} |r_\uparrow ^{nm}|^2 \right ]
\end{equation}

for the spin up conductance and for the mixing conductance $G^{\uparrow \downarrow}$

\begin{equation}
G^{\uparrow \downarrow} = \frac{e^2}{h}\left [M - \sum_{nm} r_\uparrow ^{nm} (r_\downarrow ^{nm})^* \right ]
\end{equation}

where $M$ the total number of modes. The spin mixing conductance affects only the component of the spin accumulation perpendicular to the the electrode's magnetisation by rotating the spins around it, see the last two terms of eq.~(\ref{eq:spin_current}). The validity of the above expressions is restricted to the case in which the contacts limit the total conductance\cite{bauer_2003}.

In our experiment, we use FM electrodes both for the injection of a spin polarised current and for the detection of the spin accumulation. We now derive from the circuit theory some relationships relevant in the two cases: FM electrode as voltage probe and FM electrode as spin source.

When a FM electrode is used as a voltage probe, the spin independent chemical potential on the FM side is raised above the N chemical potential by an amount that depends on the spin accumulation on the two sides of the contact. To see this, we set $i = 0$ in eq.~(\ref{eq:charge_current}) and obtain

\begin{equation}
\mu^F_j = \mu + P_j \left (\mathbf{m}_j \cdot \bm{\mu} - |\bm{\mu}^F_j|\right ).
\label{voltage_electrode}
\end{equation}

$P_j$ being the ``efficiency'' of the detector.

When the FM electrode is used as a spin injector, the charge current carries along a spin current. We use eq.~(\ref{eq:charge_current}) in eq.~(\ref{eq:spin_current}), as we control in our experiment the charge current, and we find and expression relating $\mathbf{i}_j(x)$ and $i_j(x)$:

\begin{align}
e^2 \mathbf{i}_j(x) =&
P_j e^2 i_j\mathbf{m}_j + G_j (1 - P^2_j) \left (\bm{\mu}^F_j - (\mathbf{m}_j\cdot \bm{\mu}) \mathbf{m}_j\right )  \nonumber \\ 
- 2ImG^{\uparrow \downarrow}_j& \mathbf{m}_j \times \bm{\mu} + 2ReG^{\uparrow \downarrow}_j \mathbf{m}_j \times (\mathbf{m}_j \times \bm{\mu}).
\label{spin_curr}
\end{align}

The first term shows that a charge current $I = ei$ carries a spin current $\mathbf{I} = \mu_B \mathbf{i}$, $\mu_B$ being the Bohr's magnetone, with an efficiency given by the polarisation $P$ of the interface. The second term describes a decrease in conductance because one spin channel is partially blocked.

If the contacts have much higher resistance than the ferromagnetic region in which spin accumulation occurs, the particle currents are thus determined by the large voltage drop across the interface and the small spin accumulation in the ferromagnet can be neglected altogether, $\bm{\mu}^F = 0$. This is valid in the limit $G \ll \sigma_F \lambda^{-1}_F$ (in the order of Ohms for thin FM layers), where $\sigma_F$ and $\lambda_F\approx 50$~nm are respectively the conductivity and the spin relaxation length of the ferromagnet.

For the spin independent chemical potential, one has to solve the diffusion equation 

\begin{equation}
-D\nabla ^2\mu = 0
\end{equation}

with the boundary condition set by the charge current (its direction is along the gradient of the chemical potential)

\begin{equation}
i_j(x) = \nu_D D |\bm{\nabla} \mu|
\end{equation}

For the spin dependent chemical potential, in the limit $\tau_{dif\!f} \ll \tau_{sf}$, one can neglect the diffusion term $-D\nabla ^2\bm{\mu}$ and assume a uniform spin accumulation $\bm{\mu}_a$ throughout the island. In the steady state, the injection of magnetisation has to compensate for the relaxation:

\begin{equation}
\frac{1}{\nu_D\hat{V}} \sum_j \mathbf{i}_j = \frac{\bm{\mu}_a}{\tau_{sf}}
+ \frac{g\mu_B}{\hbar} \mathbf{B} \times \bm{\mu}_a
\label{dif_boundary}
\end{equation}

where the first term on the right hand side describes spin relaxation and the second spin precession in a uniform external magnetic field\cite{huertas}. Using the expression for the spin current eq.~(\ref{spin_curr}), we rearrange the terms, to show that the presence of the contacts introduces an extra mechanism for spin relaxation.

The spin dependent chemical potential can be written in the following form\cite{bauer}:

\begin{equation}
\mathrm{T}\bm{\mu}_a =
\frac{1}{\nu_D \hat{V}}\sum_j P_j i_j \mathbf{m}_j \equiv \mathbf{v}
\label{TDspin}
\end{equation}

where the term on the right hand side is the source term $\mathbf{v}$ and $\mathrm{T}$ is a $3\times3$ matrix operator

\begin{align}
\mathrm{T}\bm{\mu}_a =& \left(\frac{1}{\tau_{sf}} + \sum_j \frac{G_j(1 - P^2_j)}{\nu_D e^2 \hat{V}}\right) \bm{\mu}_a \nonumber \\
+& \left( \frac{g\mu_B}{\hbar}\mathbf{B} + \sum_j \frac{2\mathrm{Im}G^{\uparrow\downarrow}_j}{\nu_D e^2 \hat{V}}\mathbf{m}_j \right) \times \bm{\mu}_a \nonumber \\
-& \sum_j\frac{2\mathrm{Re}G^{\uparrow \downarrow}_j - G_j(1 - P_j^2)}{\nu_D e^2 \hat{V}}
 \mathbf{m}_j \times (\mathbf{m}_j \times \bm{\mu}_a)
\label{Tmatrix}
\end{align}

An explanation of the above now follows. The first term proportional to $\bm{\mu}_a$ relaxes the magnetisation via two different mechanisms: (a) the interaction of the spin with the normal metal (spin orbit scattering), occurring at a rate $\tau^{-1}_{sf}$, and (b) the leaking of the spins to the leads, proportional to the interfaces' conductance, $G_j$. The time associated with the latter is the spin escape time $\tau_{esc} \equiv \sum_j G_j (1 - P_j^2)/\nu_D e^2\hat{V}$. The total spin relaxation time is the sum of the two contributions: $\tau^{-1}_{rel} = \tau^{-1}_{sf} + \tau^{-1}_{esc}$.

This timescale $\tau_{esc}$ is relevant in a two-terminal GMR-type of measurement: a spin dependent resistance appears if electrons cross the second FM/N interface while still retaining the information about the magnetisation of the first FM/N interface. This is equivalent to having $\tau_{sf} > \tau_{esc}$.

The second term plays the role of an effective magnetic field $\bm{\omega} = g\mu_B\mathbf{B}/\hbar + \bm{\omega}_{mix}$: the magnetisation perpendicular to $\bm{\omega}$ precesses with constant Larmor frequency $|\bm{\omega}|/2\pi$ around it. The presence of the leads introduces an extra term $\bm{\omega}_{mix}$, that depends on the orientation of the contacts and changes sign if all the magnetisations $\mathbf{m}_j$ are reversed.

The last term affects only the spin accumulation perpendicular to $\mathbf{m}_j$. If this is the case, it simply becomes proportional to $\bm{\mu}_a$ and it adds up to the spin relaxation. With other words, it is responsible for the anisotropic relaxation of the magnetisation.

It was shown\cite{brataas} that the coefficients $c_j \equiv \left (2\mathrm{Re}G^{\uparrow \downarrow}_j - G_j(1 - P_j^2)\right )/\nu_D e^2 \hat{V}$ are larger than 0, resulting in an enhancement of the relaxation of the component perpendicular to the electrodes' magnetisation. This follows from the assumption that in the FM, the spin accumulation is only collinear to the electrode's magnetisation. $c_j^{-1}$ has the units of time and represents the spin mixing time $\tau_{mix}$.

Eq.~(\ref{TDspin}) can be solved by inverting the matrix $\mathrm{T}$, giving for $\bm{\mu}_a$

\begin{widetext}
\begin{equation}
\bm{\mu}_a =
\frac{a^2\mathbf{v}
+ \left( \bm{\omega} \cdot \mathbf{v} \right )\bm{\omega}
- a\bm{\omega} \times \mathbf{v}
+ \sum_i c_i(\mathbf{m}_i \cdot \bm{\omega}) \mathbf{m}_i \times \mathbf{v}
- a\sum_i c_i \mathbf{m}_i \times (\mathbf{m}_i \times \mathbf{v})
+ \frac{1}{2} \sum_{i,j} c_ic_j [(\mathbf{m}_i \times \mathbf{m}_j) \cdot \mathbf{v}] (\mathbf{m}_i \times \mathbf{m}_j)}
{a^3 + a|\bm{\omega}|^2 
- \sum_i c_i [a^2 + (\mathbf{m}_i \cdot \bm{\omega})^2]
+\frac{a}{2}\sum_{i,j} c_i c_j |\mathbf{m}_i \times \mathbf{m}_j|^2
- \frac{1}{6}\sum_{i,j,k}c_i c_j c_k 
|\mathbf{m}_i \cdot (\mathbf{m}_j \times \mathbf{m}_k)|^2}
\label{eq:prec_2D_compl}
\end{equation}
\end{widetext}

with $a \equiv \tau^{-1}_{rel} - \sum c_j$.

\begin{figure}
\includegraphics[width = 0.5\textwidth]{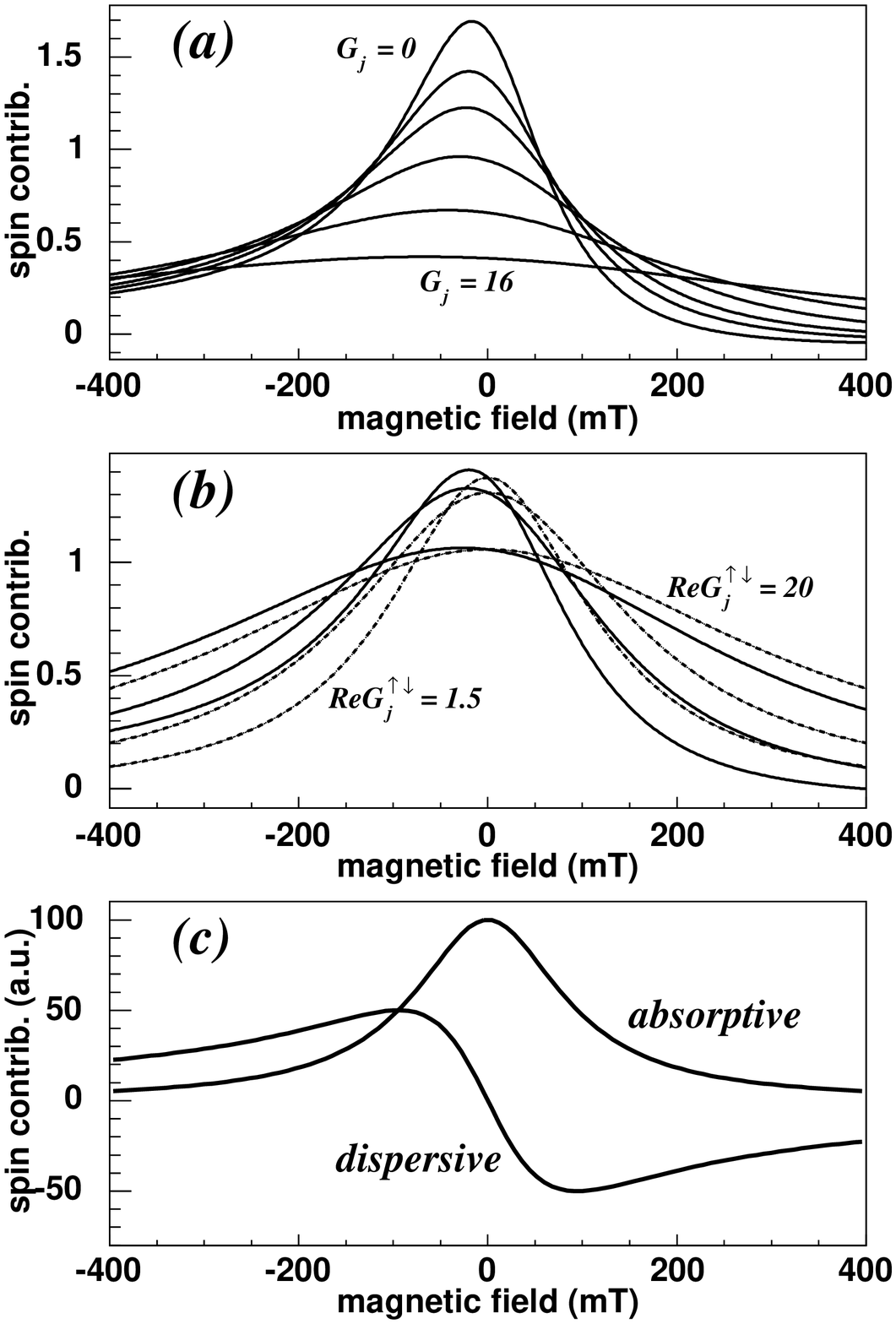}
\begin{picture}(0, 0)
\put(70, 315){\includegraphics[width = 1.3cm, height = 1.75cm]{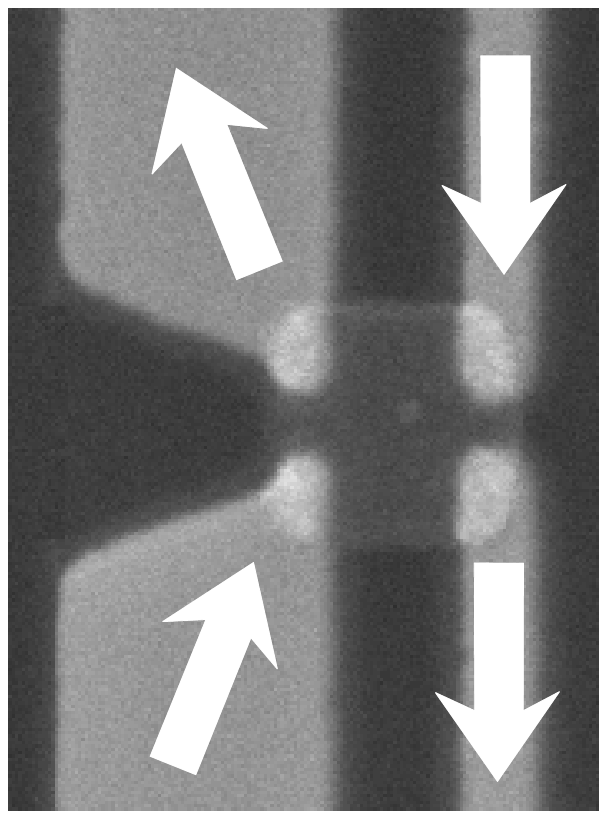}}
\put(65, 310){\vector(1, 0){40}}
\put(100, 301){$\mathbf{\hat{x}}$}
\put(65, 310){\vector(0, 1){40}}
\put(58, 340){$\mathbf{\hat{y}}$}
\end{picture}
\caption{Calculated spin signal contribution, proportional to $\bm{\mu}\ \mathbf{\cdot\ d}$ as a function of the applied magnetic field  $|\mathbf{B}|$ for a magnetic configuration represented in the inset of the top panel, with the largest electrodes tilted inwards by 20$^o$. In (a), $g_j$ are assumed to be the same for all junctions, and equal 0, 1, 2, 4, 8 and 16 in units of $\nu_D e^2 \hat{V}$ (from top curve to bottom). For all curves $P = 0.5$ and $\tau_{sf} = 60$~ps. For the mixing conductance, we have arbitrarily chosen $2g^{\uparrow \downarrow} = (1.5 + i0.1)g$. (b) shows the calculated signal for a fixed value of $g = 1$, now varying the mixing conductance $2g^{\uparrow \downarrow} = (1.5 + i0.1)k$, for k = 1.5, 5 and 20. Solid lines correspond to the field applied in the $\mathbf{z}$ direction, dashed lines for the field applied along $\mathbf{x}$. (c) The ``absorptive'' and ``dispersive'' components\cite{johnson_88} as defined by eq.~(\ref{absorptive}) and eq.~(\ref{dispersive}): in precession measurements, the detected spin signal can be written as a linear combination of these two terms, with a possible shift in the magnetic field.}
\label{sim_abs_disp}
\end{figure}

In the experiments reported here, we use precession measurements as a tool to study the spin accumulation, by applying a uniform magnetic field to the island, for instance in the $\mathbf{\hat{z}}$ direction. We now plot the dependence of the detected spin related contribution $e\Delta V = \mu^F_j - \mu^F_{j'} = \bm{\mu} \cdot \mathbf{d}$ with $\mathbf{d} = P_j\mathbf{m}_j - P_{j'}\mathbf{m}_{j'}$ (scaled by $(\nu_D e \hat{V})^{-1}$) as a function of the external magnetic field $\mathbf{B} = |\mathbf{B}| \hat{\mathbf{z}}$, for a four-contact device and the magnetic configuration depicted in the inset of Fig.~\ref{sim_abs_disp}(a) (with the magnetisation of the largest electrodes pointing inwards by 20$^o$ and lying completely on the substrate). We use renormalised parameters $g = G / \nu e^2\hat{V}$, $g^{\uparrow \downarrow} = G^{\uparrow \downarrow} / \nu e^2\hat{V}$ and we assume $P = 50\%$ and $\tau_{sf} = 60$~ps.

To show the dependence of eq.~(\ref{eq:prec_2D_compl}) on $g^{\uparrow \downarrow}$, we plot the detected signal for arbitrarily chosen values of the parameters $g^{\uparrow \downarrow}$ and $g$. Figure~\ref{sim_abs_disp}(a) shows the magnetic contribution to the total signal using $g^{\uparrow \downarrow} = (1.5 + i0.1)g$, for different values of the interface conductance $g$. The width of the curves increases from the top trace ($g = 0$) to the bottom one ($g = 16$), reflecting the fact that with increasing interface conductance, more relaxation takes place in the leads. Fig.~\ref{sim_abs_disp}(b) shows the signal for fixed $g = 1$ and different mixing conductances\cite{limits} $2g^{\uparrow \downarrow} = (1.5 + i0.1)k$, with $k$ taking the values 1.5, 5 and 20. Here again the traces broaden, but now the relaxation of the spin is truly due to the mixing term $2g^{\uparrow \downarrow}$, the third term in eq.~(\ref{Tmatrix}). The solid lines represent the precession field in the $\mathbf{z}$ direction and dashed lines for the field along $\mathbf{x}$. We also note that the maximum of the solid curves shifts to negative fields as a result of the intrinsic precession field proportional to $2\mathrm{g^{\uparrow \downarrow}}$. This is not the case for the dashed traces, because the direction of the contacts' magnetisation leads to total cancellation of the intrinsic precession field of the $x$ component.

The detected precession signal, also in the presence of the mixing conductance terms, can always be expressed as the sum of an ``absorptive'' (even) term 

\begin{equation}
\frac{1}{\tilde{\tau}_{rel}^{-2} + \omega_z^2}
\label{absorptive}
\end{equation}

and a ``dispersive'' (odd) one\cite{johnson_88}

\begin{equation}
\frac{-\omega_z}{\tilde{\tau}_{rel}^{-2} + \omega_z^2}
\label{dispersive}
\end{equation}

with $\omega_z \equiv b + \delta \omega_z$ and a suitable choice of $\delta \omega_z$ and $\tilde{\tau}_{rel}^{-2}$. Figure~\ref{sim_abs_disp}(c) shows the absorptive and dispersive terms, plotted for $\delta \omega_z = 0$.

\section{SAMPLE FABRICATION}
\label{sample_fabrication}

The system under study consists of an aluminium island with lateral dimensions of 400~nm~$\times$400~nm$\times$30~nm. Four cobalt electrodes of different width are connected to the island through tunnel barriers. A typical device is shown in Fig.~\ref{island}. Devices are fabricated by suspended shadow mask technique and by electron beam (e-beam) lithography. We begin with a trilayer consisting of 1.6~$\mu$m copolymer PMMA/MMA, 40~nm germanium and 200~nm PMMA deposited in this order on a 500~nm thermally oxidised Si substrate. After e-beam exposure and development of the trilayer\cite{jedema_02}, the germanium mask is suspended $1.6~\mu$m above the substrate. Then 30~nm Al are deposited under an angle to form the island, in an e-beam evaporation system with base pressure of $10^{-6}$~mbar. We notice that changing the evaporation rate from 0.1~nm/sec to 0.3~nm/sec reduces the Al resistivity by a factor of 2. For aluminium deposited at 0.3~nm/sec, $\rho_{300K}/\rho_{4K} = 2$, and for 0.2~nm/sec, $\rho_{300K}/\rho_{4K} = 1.3$. In the following, all the devices have been deposited at a rate of 0.3~nm/sec, unless indicated otherwise. Next, we oxidise the Al in 0.02--0.2~mbar pure oxygen for 2--5~min to produce tunnel barriers (20--500~$\Omega\mu\mathrm{m}^2$). The devices are produced with different tunnel barrier transparencies (from 1~k$\Omega$ to 40~k$\Omega$). Co leads (40~nm thick) are subsequently deposited under a different angle.

The devices are fabricated with decreasing tunnel barrier resistance to determine the lowest transparency for which the tunnel barriers still retain a sizeable spin selectivity. We also started off with the idea of measuring the mixing conductance term. In order to measure $G^{\uparrow \downarrow}$, $\tau_{mix}$ or $|\bm{\omega}_{mix}|^{-1}$ have to be comparable to the relaxation time $\tau_{rel}$. It was shown\cite{brataas} that $2\mathrm{Re} G^{\uparrow \downarrow} - G \geq 0$, the equality holding true for tunnel barriers. Also, the imaginary part $\mathrm{Im}G^{\uparrow \downarrow}$ for tunnel barriers, is of the same order of magnitude as $G$. The devices we fabricated with the highest transparencies, have tunnel barriers of $G^{-1} = 1~\mathrm{k}\Omega$ and show a spin relaxation time of $\tau_{sf} = 60$~ps and an escape time of $\tau_{esc} = 10^3\tau_{sf}$: such a system is unsuitable for a measurement of $G^{\uparrow \downarrow}$. 

Tunnel barriers with resistances 3 orders of magnitude lower could not be fabricated (and probably can not) in aluminium oxide. The alternative would be to decrease the island volume by a factor 1000, but this is not feasible with this fabrication technology.

\section{SPIN INJECTION EXPERIMENTS}
\label{this_experiment}

We first review and simplify the expressions that are relevant for our devices.

Spin accumulation measurements are done in a four-terminal geometry: we drive a current $I$ into and out of two electrodes and we detect the voltage $V$ using the other two electrodes. For the devices with the most transparent tunnel barriers we could fabricate, the mixing term accounts for a correction to the spin accumulation of $10^{-3}$. For this reason, we set $G_j = G_j^{\uparrow \downarrow} = 0$ in eq.~(\ref{eq:prec_2D_compl}). The following equations are derived from it, after some algebraic manipulation. The spin dependent contribution to the total signal $eV_s = \bm{\mu} \cdot \mathbf{d}$:

\begin{equation}
R_s = \frac{V_s}{I} = \frac{\tau_{sf}}{\nu_D e^2 \hat{V}} \mathbf{s \cdot d}
\label{eq:spin}
\end{equation} 

in the absence of magnetic field $\bm{\omega} = 0$, where $\mathbf{s} = P_1 \mathbf{m}_1 - P_2 \mathbf{m}_2$ is the source term and $\mathbf{d} = P_3 \mathbf{m}_3 - P_4 \mathbf{m}_4$ the detector, if current is sent from Co1 to Co2 and voltage is detected between Co3 and Co4. $\nu_D = 2.4 \times 10^{28}~\mathrm{eV}^{-1}\mathrm{m}^{-3}$ is the aluminium density of states at the Fermi energy and $\hat{V}$ the volume of the island.

In the presence of a magnetic field, the magnetisation perpendicular to the field precesses. When using ferromagnetic electrodes, one has also to be concerned with the stray fields generated by the electrodes themselves, that thread the island. To account for these, we add an extra term $\bm{\omega}_{st}$ to the external magnetic field $g\mu_B|\mathbf{B}|\mathbf{\hat{z}} /\hbar = \mathbf{\hat{z}}$, $\bm{\omega} = b \hat{\mathbf{z}} + \bm{\omega}_{st}$.

In general

\begin{eqnarray}
R_s = \frac{\tau_{sf}}{\nu_D e^2 \hat{V}} \Big[ s_{||}\cdot d_{||} + 
\frac{\mathbf{s_\mathrm{\perp} \cdot d}_\mathrm{\perp} - (\mathbf{s}_\perp \times \mathbf{d}_\perp) \mathbf{\cdot} \bm{\omega} \tau_{sf}}{1 + |\mathbf{\omega}|^2 \tau_{sf}^2} \Big]
\label{eq:prec_2D}
\end{eqnarray}

where $||$ means in the same direction as the external field. This is the $\mathbf{z}$ direction if also $\bm{\omega}_{st}$ has only a $\mathbf{z}$ component, $\omega_{st,z}$. As mentioned before, eq.~(\ref{eq:prec_2D}) can be written as a linear combination of the absorptive and dispersive terms (and a constant term), eq.~(\ref{absorptive}) and eq.~(\ref{dispersive}), shifted in the precession field by $\omega_{st, z}$, with $\hat{\tau}_{rel}^{-2} = \tau_{sf}^{-2} + \omega_{st, \perp}^{2}$.

Eq.~(\ref{eq:prec_2D}) is not invariant under the reversal of the electrodes' magnetisation $\mathbf{s} \rightarrow -\mathbf{s}$ and $\mathbf{d} \rightarrow -\mathbf{d}$, because the reversal also produces a change of sign $\bm{\omega}_{st} \rightarrow -\bm{\omega}_{st} \neq \mathbf{0}$. It obeys, however, the reciprocity relation\cite{buttiker} that requires the interchange of voltage and current probes and the reversal of all magnetic fields and magnetisations. In the following we show that we obtain experimentally the same spin signal but only in the constraints given by the reciprocity theorem.

\section{MEASUREMENT CONFIGURATIONS}
\label{meas_configuration}

\begin{figure}

\begin{picture}(0, 0)
\put(0, 0){\normalsize a)}
\end{picture}
\mbox{\parbox{0.15\textwidth}{\includegraphics[width = 2.0cm, height = 2.7cm]{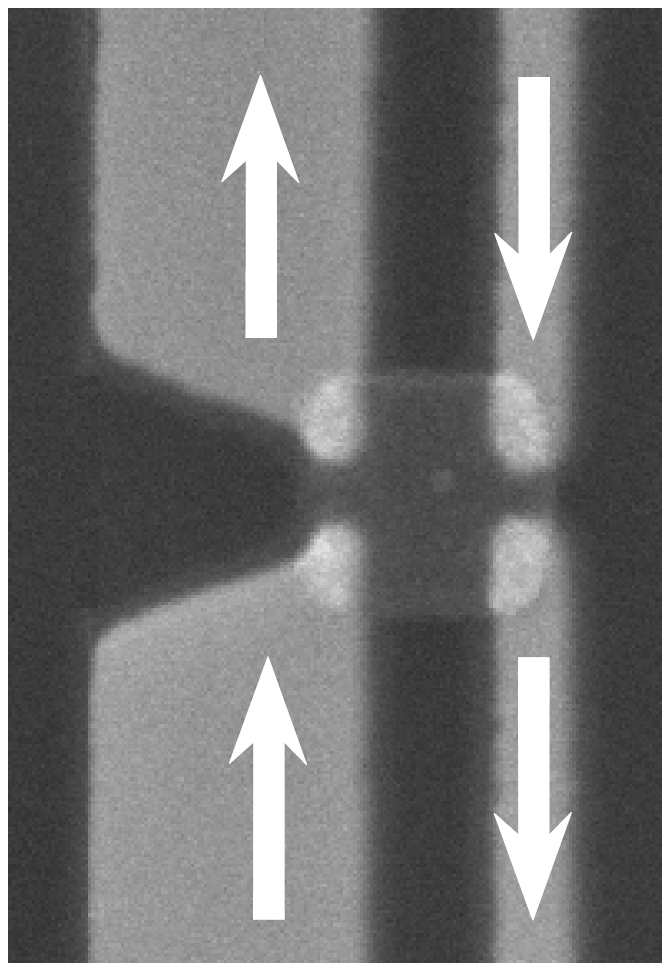}\\{\normalsize antiparallel}}
\hspace{0.0cm}
\parbox{0.15\textwidth}{\includegraphics[width = 2.0cm, height = 2.7cm]{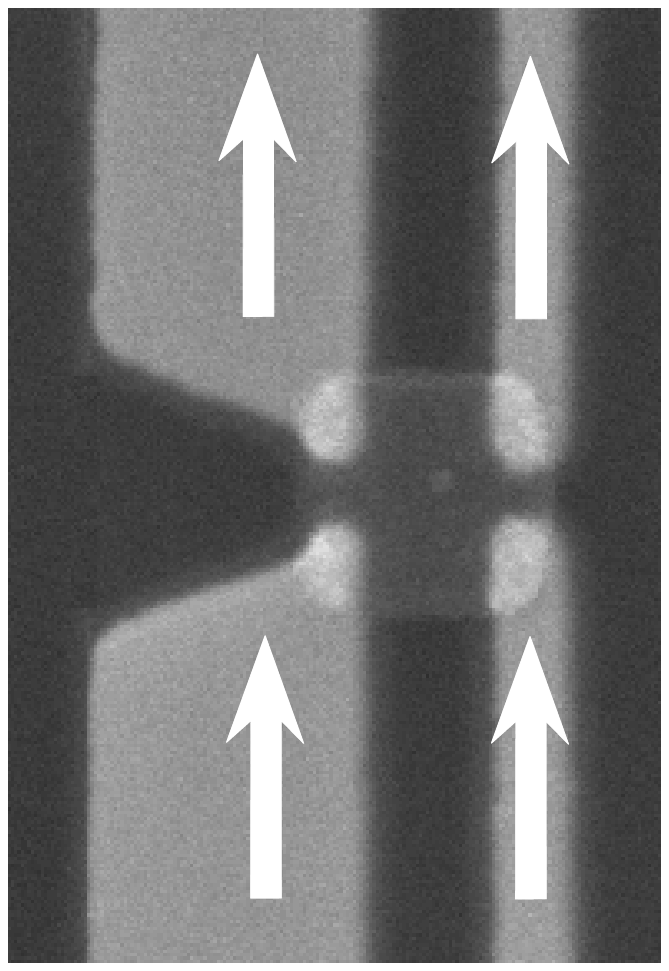}\\{\normalsize parallel}}
\hspace{0.0cm}
\parbox{0.15\textwidth}{\includegraphics[width = 2.0cm, height = 2.7cm]{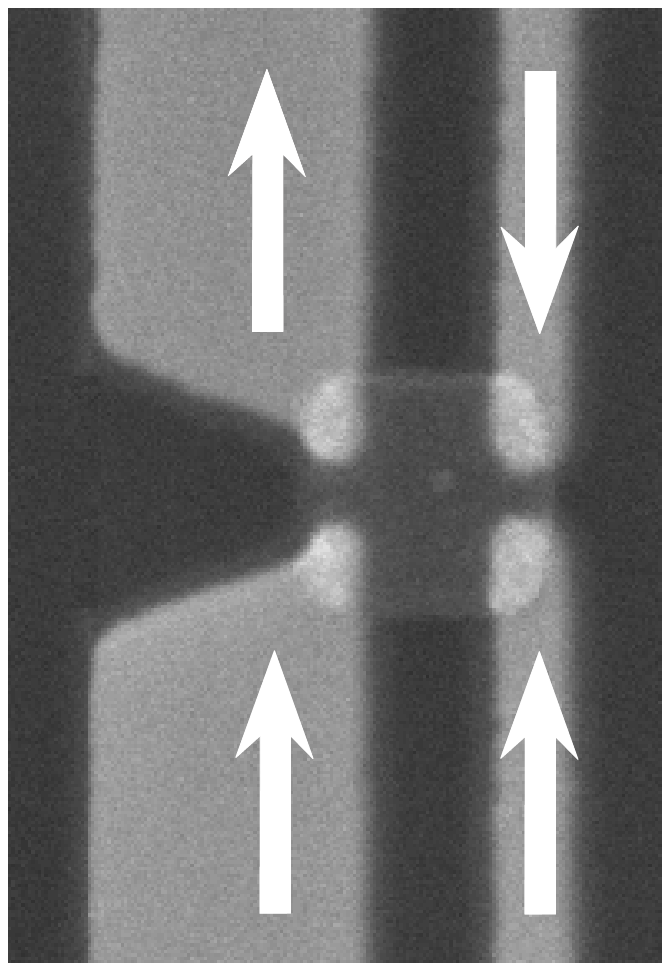}\\{\normalsize anomalous}}}

{\vspace{0.2cm}}

\begin{picture}(0, 0)
\put(0, 0){\normalsize b)}
\end{picture}
\mbox{\parbox{0.15\textwidth}{\includegraphics[width = 2.0cm, height = 2.7cm]{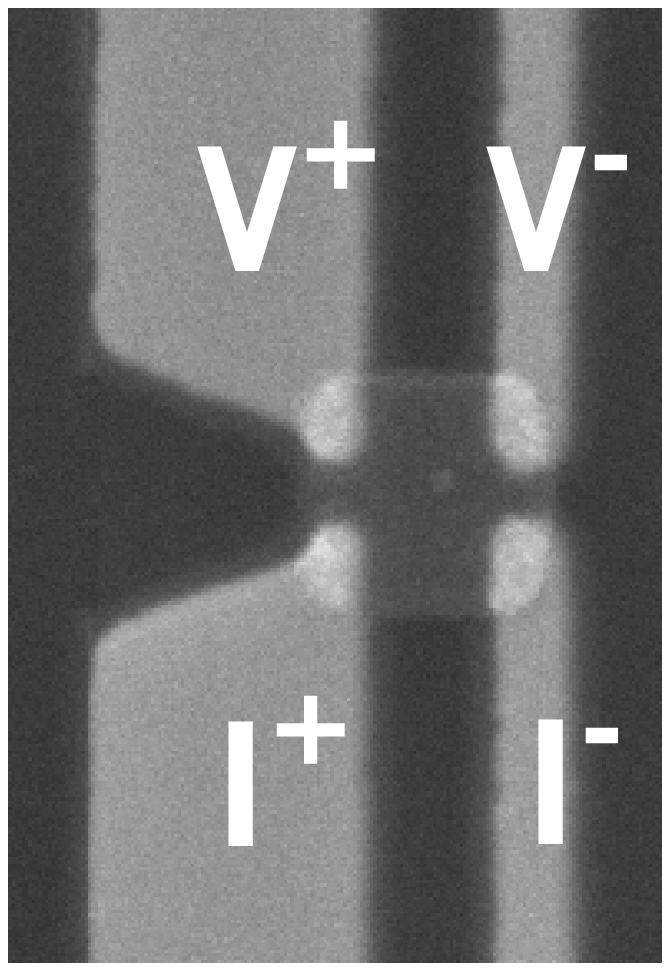}\\{\normalsize side}}
\hspace{0.0cm}
\parbox{0.15\textwidth}{\includegraphics[width = 2.0cm, height = 2.7cm]{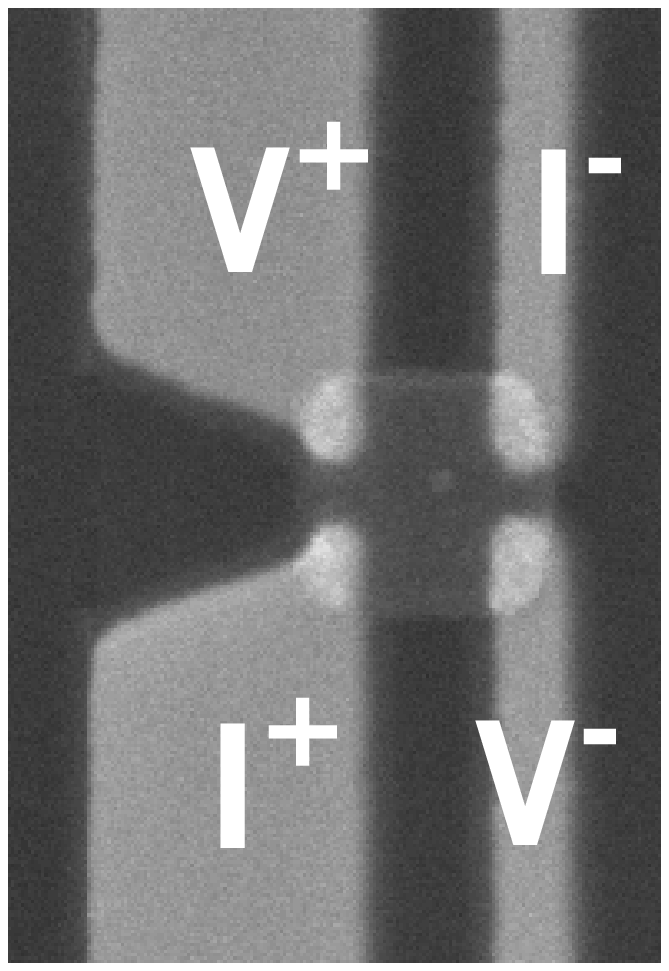}\\{\normalsize diagonal}}
\hspace{0.0cm}
\parbox{0.15\textwidth}{\includegraphics[width = 2.0cm, height = 2.7cm]{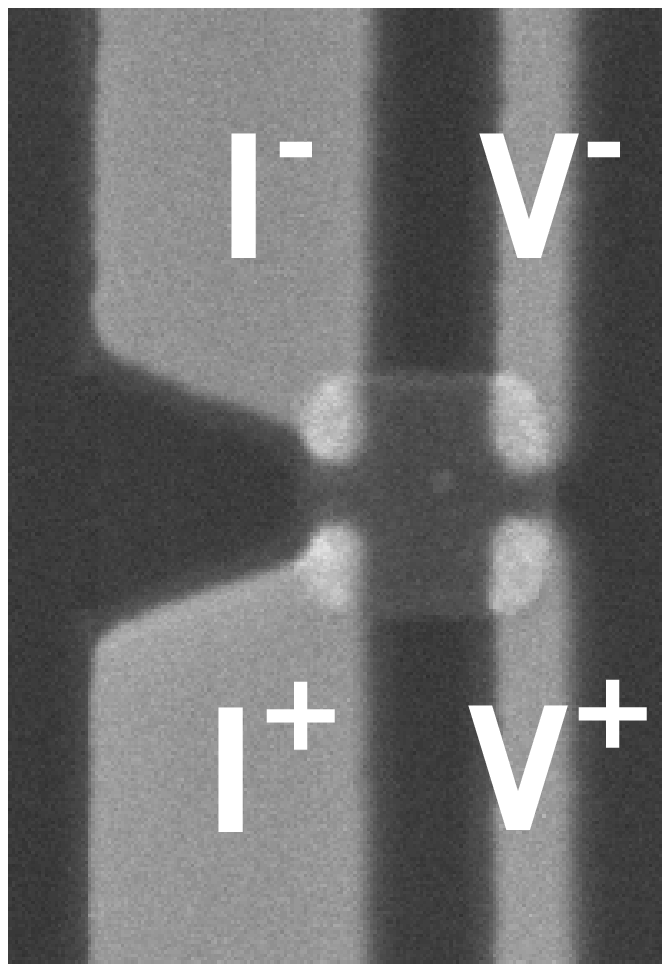}\\{\normalsize opposite}}}

\vspace{0.2cm}

\begin{picture}(0, 0)
\put(-7, 40){\normalsize c)}
\end{picture}
\includegraphics[width = 0.95\linewidth]{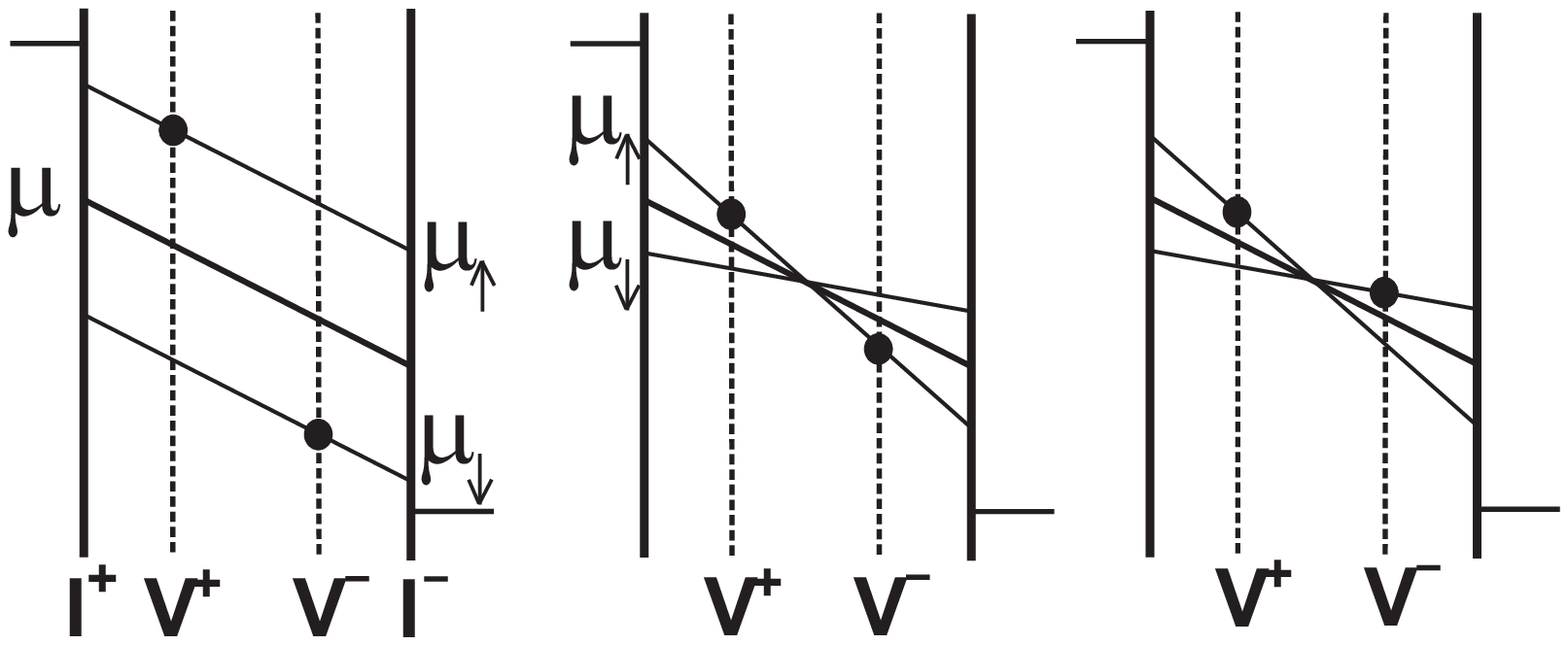}
\caption{(a) The three magnetic configurations and (b) the three possible independent measuring configurations: Current is sent from $I^+$ to $I^-$, the detected voltage is $V = V^+ - V^-$. (c) The chemical potentials $\mu_\uparrow = \mu + |\bm{\mu}|$ and $\mu_\downarrow = \mu - |\bm{\mu}|$ inside the island for the \textit{side} configuration in the \textit{antiparallel}, \textit{parallel} and \textit{anomalous} cases, assuming collinear magnetisation. We remind that only for antiparallel injectors, is $|\bm{\mu}|$ uniform (see main text). The lines represent the spin up and down chemical potentials $\mu_{\uparrow,\downarrow}$ and the thick line the average $\mu$. The black dots indicate the potential measured by the $V^+$ and $V^-$ probes for $P = 1$.}
\label{setup}
\end{figure}

The four cobalt electrodes have different widths, one pair 100~nm wide, the other 500~nm, with the latter having the lowest coercive field. The magnetic shape anisotropy holds the electrodes' magnetisation in the substrate plane, and by applying an in-plane external magnetic field along the electrodes' direction (the $\bf{y}$ direction), we can independently reverse the direction of the magnetisation of the electrodes. We identify an \textit{antiparallel} configuration, in which two electrodes are pointing in the same direction and two in the opposite, and a \textit{parallel}, in which all four have the same direction, as shown in Fig.~\ref{setup}(a). Here \textit{parallel} and \textit{antiparallel} are used as a practical shorthand notation: we will show in fact that the electrodes' magnetisations are non-collinear. In the \textit{anomalous} configuration, three electrodes (the two wide and one of the narrow ones) are pointing in the same direction and the fourth narrow electrode is in the opposite direction.

Figure~\ref{setup}(b) shows the three independent electrical measuring configurations. The current $I$ is sent between $I^+$ and $I^-$ and the detected voltage is $V = V^+ - V^-$. The plotted signal is $R = V / I$. In the \textit{side} configuration, the background signal is the island's Ohmic resistance. In the \textit{diagonal} configuration, little Ohmic contribution is expected, owing to the symmetric position of the voltage contacts with respect to the current path. The spin dependent contribution in the two cases is however equal if the island is zero-dimensional. On the other hand, the \textit{opposite} configuration should show small spin signal as the widest electrodes switch at the same time ($\mathbf{s} \approx \mathbf{0}$) and so do the narrow ones ($\mathbf{d} \approx \mathbf{0}$).

We write the total signal as the sum of a spin independent (Ohmic) and a spin dependent contribution, $R = R_{Ohm} + R_s$. $R_{Ohm}$ is the island four-terminal Ohmic resistance and we assume it to be independent of the magnetic arrangement of the electrodes (we exclude for instance the Hall effect).

$R_s$ is the spin-related part. We refer to Fig.~\ref{setup}(c) to illustrate its contribution in the three different magnetic configurations. Suppose for the moment that all electrodes are collinear and the barrier polarisations $P_j$ are all equal. $I^+$ and $I^-$ are the current electrodes, $V^+$ and $V^-$ the voltage probes and the black dots are the voltages that would be detected if the polarisation was $P = 1$. The position of the dot on the $\mu_\uparrow$ or $\mu_\downarrow$ lines depends on the orientation of the detector. In the antiparallel configuration, a uniform non-equilibrium magnetisation in the island is created and $(\mu_\uparrow - \mu_\downarrow) / 2 \equiv \mu_y \neq 0$. This potential difference is detected at the voltage electrodes and the signal $R_s$ is given by eq.~(\ref{eq:prec_2D}).

In the parallel configuration, there is no net spin accumulation. However, a spin \textit{current} $|\mathbf{I}| = PI\mu_B/e$ is injected at $I^+$ and extracted at $I^-$, giving rise to a space dependent magnetisation, $|\mathbf{I}| = - (\sigma_N \mu_B / e) \cdot \nabla (\mu_\uparrow - \mu_\downarrow) / 2$, $\sigma_N$ being the Ohmic conductance of the island in much the same way a charge current generates a space dependent chemical potential, $I = -(\sigma_N / e) \nabla \mu$. Recalling that the device is in the parallel configuration, the detected spin related contribution is $P$ times the difference of the spin up chemical potential at the $V^+$ and $V^-$ positions. The spin signal is a fraction of the Ohmic resistance, $R_s = P^2R_{Ohm}$. For $P = 1$, the total resistance doubles because only one spin channel is used, halving the island's conductance.

In the anomalous configuration, the $V^+$ probe measures $\mu_\uparrow$ and $V^-$ detects $\mu_\downarrow$. Owing to the symmetric position of $V^+$ and $V^-$ with respect to $I^+$ and $I^-$, both probes measure the same amount of magnetisation and the spin dependent contribution is cancelled\cite{anomalous_equivalent}, $R_s = 0$. In the anomalous configuration, one therefore expects to have the lowest signal, equal to the island's Ohmic resistance.

Standard lock-in techniques are employed, with excitation currents ranging from 5~$\mu$A to 100~$\mu$A and with modulation frequencies between 4--10~Hz.

\section{EXPERIMENTAL RESULTS}
\label{results}

A spin valve experiment is a four terminal resistance  measurement in one of the three possible configurations, see Fig.~\ref{setup}(b), as a function of the in-plane magnetic field (in the $\mathbf{y}$ direction). Spin valve measurements were performed both at 4.2~K and at RT, and precession measurements only at RT. We measured 9 devices in total, one at 4.2~K only, two both at RT and 4.2~K, and 6 only at RT. They all show consistent behaviour.

We report here a complete set of measurements on a device with tunnel barriers of $R_1 = 1.5~\rm{k}\Omega$, $R_2 = 0.90~\rm{k}\Omega$, $R_3 = 1.6~\rm{k}\Omega$ and $R_4 = 0.75~\rm{k}\Omega$ at RT\cite{TBresistances}. From these, we show that we can derive the magnetisation orientation of the magnetic contacts and their polarisations. Comparison measurements on a different device at 4.2~K are shown at the end of this section. We will also discuss the relationship between tunnel barriers' transparencies and polarisation.

To characterise the device, each individual tunnel barrier is first measured: with reference to Fig.~\ref{island}, by sending a current between Co1 and Co2 and detecting the voltage between Co1 and Co3, we measure tunnel barrier 1. Usually Co1 and Co3 have the same resistance (within 20\%) as do Co2 and Co4, because they have nominally the same area.

\begin{figure}
\includegraphics[width = 0.45\textwidth, height = 0.25\textwidth]{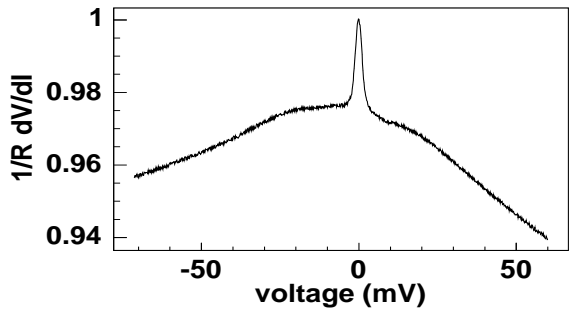}
\caption{Normalised differential resistance of a 1.8~k$\Omega$ tunnel barrier at 4.2~K as function of the dc voltage: a peak appears at zero-bias and the curve shows an asymmetry. Positive voltage means current flowing from Co to Al.}
\label{TB}
\end{figure}

The I--V characteristic of a 1.8~k$\Omega$ tunnel barrier measured at liquid helium temperature is shown Fig.~\ref{TB}. Positive voltage means Co at higher potential than Al. All the tunnel barriers (TB) we measure at 4.2~K (with resistances down to 7~k$\Omega$) show a peak at zero bias and are asymmetric in the applied bias. Variation of the tunnel barrier differential resistance of 10\% in the bias range used is visible.

\begin{figure}
\includegraphics[width = 0.45\textwidth]{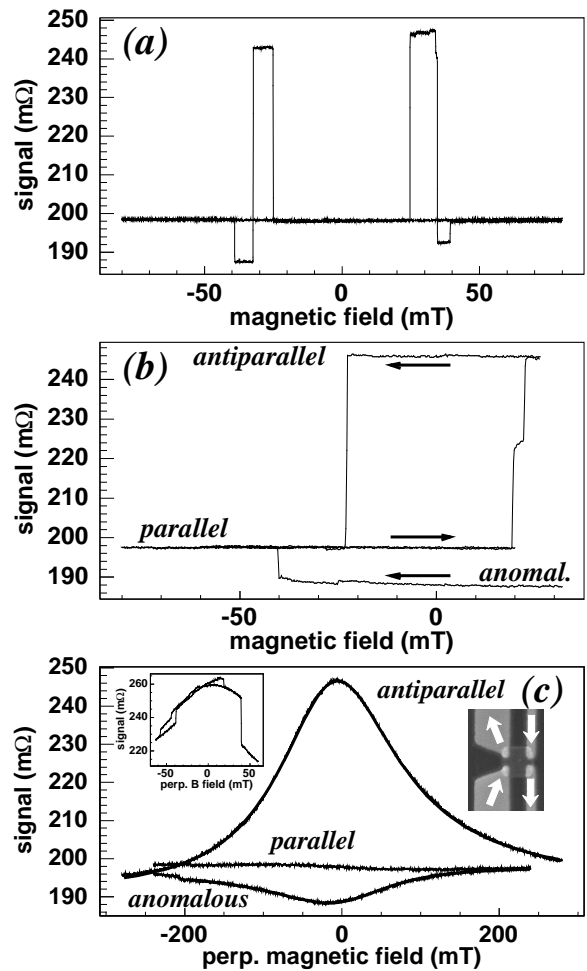}
\begin{picture}(0, 0)
\put(-55, 65){\includegraphics[width = 1.0cm, height = 1.35cm]{antiparallel_tilted}}
\end{picture}
\caption{(a) Spin valve measurement of $R = (V^+ - V^-)/ I$ as a function of the in-plane magnetic field at RT. (b) \textit{Memory effect} starting from the antiparallel configuration (upper trace beginning at 246~m$\Omega$) and from the anomalous configuration (lower trace from 186~m$\Omega$). (c) Precession measurements as a function of the magnetic field applied perpendicular to the substrate for the antiparallel, parallel and anomalous configurations. The fits for the antiparallel and anomalous cases follow closely the experimental data. The right inset shows the direction of the electrodes' magnetisations. The left inset shows precession measurements with the external field applied perpendicular to the spin accumulation in the $\mathbf{x}$ direction. At fields as low as 30~mT, the contacts' domains begin to turn, and the signal becomes irregular.}
\label{g_side}
\end{figure}

The measurements are organised in the following way: for each measuring configuration, we show one spin valve measurement and three precession measurements for the different magnetic configurations.
 
Figure~\ref{g_side}(a) shows measurements at RT for the side configuration. Starting with the magnetic field at $80$~mT, with all the magnetic contacts pointing parallel to each other, we sweep the field to negative values. At $-25$~mT, the two larger electrodes flip, the magnetic configuration is antiparallel and the detected signal increases above the background level by 45~m$\Omega$. Increasing the field further, at $-32$~mT, one of the smaller electrodes reverses, and the signal dips 10~m$\Omega$ below the background level. At $-38$~mT, the second smallest electrode also flips and the signal reaches the background value. The reverse trace shows very similar behaviour, the notable difference being a larger peak, about 48~m$\Omega$, and a smaller dip, 6~m$\Omega$. Repeated sweeps give similar results.

Figure~\ref{g_side}(b) shows the \textit{memory effect}, that reflects the hysteretic behaviour of the electrodes. Starting with the system in the antiparallel configuration at $+30$~mT, we sweep the field toward negative magnetic fields. The electrodes stay in the antiparallel magnetic configuration until we reach $-25$~mT, at which point the largest electrodes switch parallel to the smallest ones. In the reverse sweep, at $+20$~mT the largest electrodes flip again, returning the initial configuration. The second trace is the memory effect in the anomalous configuration. Suppose Co1, Co2 and Co3 are parallel to the $\mathbf{y}$ direction and Co4 is opposite. Starting at $+33$~mT (and 186~m$\Omega$), we sweep the field towards negative fields. At $-25$~mT, the largest electrodes Co1 and Co3 flip, now pointing in the $\mathbf{-y}$ direction and parallel to Co4: this is still an anomalous configuration. Upon reaching 40~mT, the smallest electrode Co3 flips parallel to the other three and the detected signal reaches the background level.

In a precession experiment, we apply an external magnetic field perpendicular to the sample (in the $\mathbf{z}$ direction), with the in-plane field switched off. Figure~\ref{g_side}(c) shows precession measurements for the three magnetic cases, parallel, antiparallel and anomalous configurations. In both antiparallel and anomalous there is a noticeable dependence of the signal on the magnetic field, whereas in the parallel case little modulation is seen. From the smoothness of the curve, we conclude that the contacts' domains do not flip irreversibly in the direction of the external field, up to fields of 280~mT. At higher fields, however, the magnetisation of the end domains of the strip is unstable and tends to flip to a different configuration irreversibly, resulting in sudden jumps of the signal. One could also think of doing precession measurements with an in-plane magnetic field, perpendicular to the leads, in the $\mathbf{x}$ direction. The left inset in Fig.~\ref{g_side}(c) shows the result of such a measurement, a sweep from +40~mT to --60~mT and back to +60~mT: at fields as low as $|30|$~mT, the electrodes' domains begin to rotate and the signal deviates from a smooth curve.

The signal is fit with eq.~(\ref{eq:prec_2D}), written in a way suitable for interpolation, and similar to eq.~(2) of Ref.~\onlinecite{zaffalon}:

\begin{equation}
 R = \frac{|\mathbf{s}_\perp||\mathbf{d}_\perp|\tilde{\tau}_{sf}}{e^2\nu_D \hat{V}}\frac{\cos\phi -\omega_z \tilde{\tau}_{sf}\sin\phi}{1 + \omega_z^2 \tilde{\tau}_{sf}^2} + R_{back}.
\label{eq:prec_2D_fit}
\end{equation}

The background term $R_{back}$ accounts not only for the Ohmic resistance, but also for the magnetisation that is injected parallel to the applied field and that does not precess. In fact, from the geometry of the device, the end domain of the contacts are pointing slightly upwards.

\begin{figure}
\includegraphics[width = 0.95\linewidth]{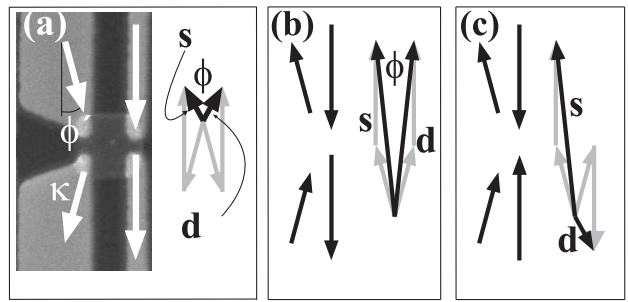}
\caption{(a) Orientation of the electrodes' magnetisation in the parallel configuration. The quantity $P\mathbf{m}_\perp$ for the widest electrodes is canted by an angle $\phi'$ with the direction of the narrow electrodes' magnetisation and is shorter by $\kappa$ than the narrowest electrodes. The black arrows represent the injector $\mathbf{s}_\perp = (P_1 \mathbf{m}_1 - P_2 \mathbf{m}_2)_\perp$ and the detector $\mathbf{d}_\perp = (P_3 \mathbf{m}_3 - P_4 \mathbf{m}_4)_\perp$, $\phi$ being the angle between the two vectors $\mathbf{s}_\perp$ and $\mathbf{d}_\perp$. The same schematics for the antiparallel (b) and anomalous (c) configurations.}
\label{arrows}
\end{figure}

Note that only in the case $\omega_{st, \perp} = 0$, $\phi$ is the angle between injector and detector (it is actually the projection of the angle on the plane perpendicular to $\mathbf{z}$), and $\tilde{\tau}_{rel} = \tau_{sf}$. Here, we allow for a stray field through the sample only in $\mathbf{z}$ direction: $\omega_z = g\mu_B |\mathbf{B}| /\hbar + \omega_{st, z}$ and we assume $\tilde{\tau}_{rel}^{-2} = \approx \tau_{sf}^{-2}$.

For the antiparallel case of Fig.~(\ref{g_side}b) we find $\tau_{sf} = 62 \pm 2$~ps  $\phi = (-0.06 \pm 0.01) \pi$, $\omega_{st} = -14 \pm 4$~mT and a background $R_{back} = 192 \pm 1$~m$\Omega$. Using the diffusion constant for the aluminium $D = 5 \times 10^{-3} \mathrm{m}^2/\mathrm{s}$ found from resistivity measurements, the diffusion time is $\tau_{dif\!f} = L^2 / D \approx 30$~ps, shorter than $\tau_{sf}$. The spin diffusion length is $\lambda_{s} = \sqrt{D\tau_{sf}} = 550$~nm.

We now give a first estimate of the polarisation $P$: assuming that only the widest electrodes' magnetisations are rotated by $\phi$, $|\mathbf{s}| = |\mathbf{d}| = 2P\cos (\phi /2)$, we find $P = 7 \%$.

Parallel and anomalous configurations differ by about $6-10~\mathrm{m}\Omega$ in the spin valve measurement, Fig~\ref{g_side}(a), and $12~\mathrm{m}\Omega$ in the precession trace, whereas the spin current accounts for only $P^2 R_{Ohm} = 1$~m$\Omega$. The precession data for the anomalous configuration indicate that accumulation also occurs\cite{error}, and accounts for most of the signal.

We fit the signal in the anomalous configuration with eq.~(\ref{eq:prec_2D}), but now fixing $\tau_{sf}$ to the value found in the (side) antiparallel case. We find $\phi = (0.11 \pm 0.04) \pi$, $\omega_{st} = 0\pm 8$~mT and $R_{back} = 197 \pm 1~\mathrm{m}\Omega$.

From the precession measurements, we work out the magnetic configuration of each electrode. We note first that for the function

\begin{equation}
g(x) = \frac{\cos \phi - x \sin \phi}{1 + x^2},
\end{equation}

$\max (g) - \min (g) = 1$ holds, for every value of $\phi$: the amplitude of the spin signal in a precession measurement is proportional to $|\mathbf{s}_\perp| |\mathbf{d}_\perp|$, independent of the angle between injector and detector.

We show now that the precession measurements for the three magnetic configurations are consistent if one assumes that the narrow electrodes Co2 and Co4 point in the $\mathbf{y}$ direction, that Co1 and Co3 are tilted inwards by an angle $\phi ' = (0.08 \pm 0.03) \pi$ and that their component on the $x-y$ plane, $|P\mathbf{m}_{\perp}|$ (i.e. the component that precesses) is smaller than the narrow ones by a factor $\kappa = 0.7 \pm 0.1$, see Fig.~\ref{arrows}. In fact, $|\mathbf{s}_\perp| |\mathbf{d}_\perp| \approx (1 + \kappa)^2$ in the antiparallel case ($\cos (\phi ') \approx 1 $) and $\approx (1 - \kappa ^2)$ for the anomalous configuration. Their ratio is $(1 - \kappa ^2) / (1 + \kappa)^2 = (0.3 / 1.7) = 18 \%$. We now show that this value is close to the experimental result. From the measurements of Fig.~\ref{g_side}(c), we find the maximum modulation of the precession signal in the anomalous and antiparallel configurations, respectively 9 and 55~$m\Omega$; their ratio being 16\%.

This is also compatible with the small magnetic field dependence of the parallel configuration signal. In fact, the ratio between magnetic signals in the parallel and antiparallel cases ($3~\mathrm{m}\Omega / 55~\mathrm{m}\Omega \approx 6\%$) is close to the expected ratio, $(1 - \kappa)^2/(1 + \kappa)^2 = (0.30 / 1.70)^2 = 3\%$ ($\cos \phi' \approx 1$). With these corrections, the efficiency of the narrowest electrodes becomes $P = 8\%$.

\begin{figure}
\includegraphics[width = 0.5\textwidth, height = 0.5\textwidth]{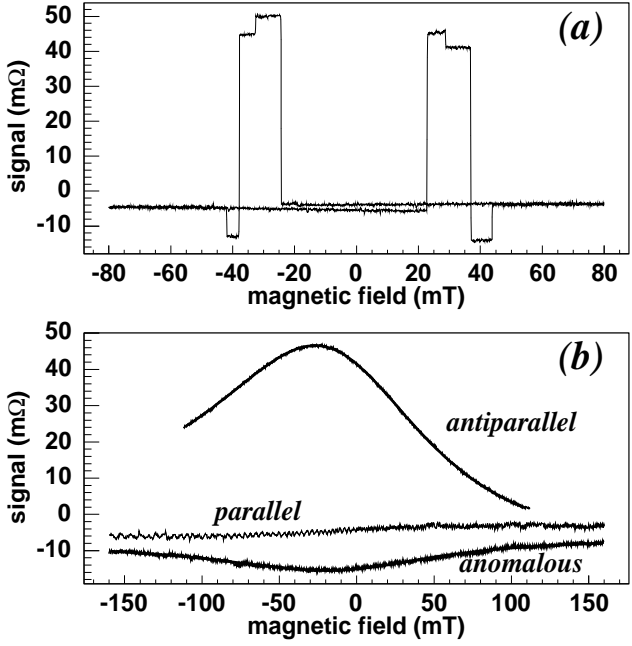}
\caption{Spin valve measurement (a) and precession measurements for the three different magnetic configurations (b) in the diagonal case. The fit for the antiparallel and anomalous configurations are superimposed onto the experimental data.}
\label{g_diagonal}
\end{figure}

Taking into account the efficiency of the detector P, the spin accumulation is $R_s/P = 850~\mathrm{m}\Omega$, larger than the Ohmic background resistance. With a typical driving current of 10~$\mu$A, we find the imbalance between up and down spins in the island $\Delta n = R_s I \nu_D e \hat{V} / P = 10^3$. For comparison, the total number of free electrons is $10^9$.

The spin signal in the side configuration is compared to that in the diagonal configuration, Fig.~\ref{g_diagonal}(a). The signal of the spin valve measurement, 54~m$\Omega$ for the left peak and 48~m$\Omega$ for the right one, is comparable to that in the side configuration, supporting the assumption of the spin accumulation in the island being uniform.

Using this measuring configuration, we also perform precession measurements in the parallel, antiparallel and anomalous magnetic configurations. As before, we find the relevant parameters, $P = (8.0 \pm 0.5)\%$ and $\tau_{sf} = 65 \pm 4$~ps, consistent with those found in the side configuration ($R_{back} = -8 \pm 1$~m$\Omega$, $\omega_{st} = -13 \pm 4$~mT and $\phi = (0.10 \pm 0.02)\pi$). In the anomalous diagonal configuration, we fix the relaxation time found in the diagonal antiparallel case and we find $\phi = 0.08 \pm 0.05$, $\omega_{st} = -8 \pm 8$~mT, $R_{back} = -7 \pm 1 $~m$\Omega$.

Both $\omega_{st}$ and $\phi$ show variations between successive measurements, spanning from $|7|$ to $|15|$~mT and from $|0.04| \pi$ to $|0.07| \pi$ respectively, in the antiparallel side configuration. On the other hand, both the spin relaxation time and the polarisation showed constant values throughout the time of the measurements.

\begin{figure}
\includegraphics[width = 0.5\textwidth, height = 0.5\textwidth]{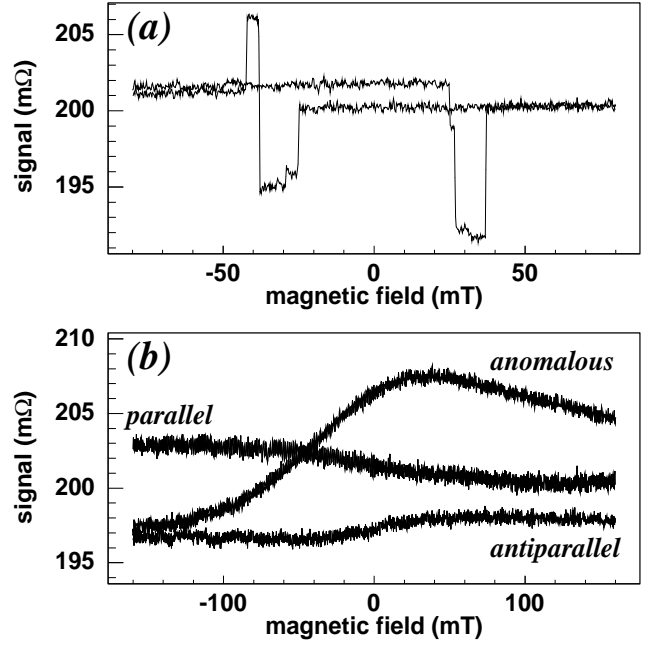}
\caption{(a) Spin valve measurement and (b) precession measurements for the three different magnetic configurations in the opposite measuring configuration. The anomalous configuration for the precession measurement is set by applying $-40$~mT in the $\mathbf{y}$ direction.}
\label{g_opposite}
\end{figure}

In the opposite configuration, the signal in the parallel and antiparallel configurations differ by $5-8~\mathrm{m}\Omega$. The precession measurements show dependence on the external B field of less than 3~m$\Omega$ and we conclude that the magnetisation is injected parallel to the external field.

The signal in the anomalous opposite configuration shows a magnetic field dependence. The most notable feature is that the signal is odd in the magnetic field, implying that $\mathbf{s}$ and $\mathbf{d}$ are almost perpendicular to each other. We fit the signal by fixing the spin relaxation time $\tau_{sf} = 62$~ps as found from the side configuration and find $\phi = (-0.35 \pm 0.01)\pi$, $\omega_{st} = -23 \pm 7$~mT and $R_{back} = 200 \pm 1$~m$\Omega$.

We now compare the predicted signal in the side antiparallel configuration, $(1 + \kappa)^2$ and in the anomalous opposite configuration $2\times 2\kappa \sin \phi '$, $(1 + \kappa)^2 / 4\kappa \sin \phi ' = 4.1$, comparable to the amplitudes ratio of the measured signal $ 55~\mathrm{m}\Omega / 10~\mathrm{m}\Omega = 5.5$.

\begin{figure}
\includegraphics[width = 0.4\textwidth, height = 0.3\textwidth]{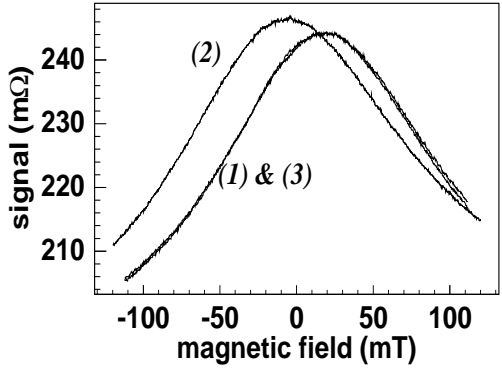}
\caption{Precession measurements in the antiparallel side configuration to show the reciprocity theorem: curve (1) is measured with magnetic field applied in the $\mathbf{z}$ direction, (2) is measured after interchange of current and voltage probes and magnetic field applied in the $-\mathbf{z}$ direction, curve (3) after reversal of all magnetisations, interchange of voltage and current probes and magnetic field in the $-\mathbf{z}$ direction.}
\label{recip}
\end{figure}

We also test the prediction of the reciprocity theorem, which states that a four-probe measurement is invariant upon exchange of the voltage and current probes and magnetic field reversal\cite{buttiker}. In the case of magnetic electrodes, one has also to reverse their magnetisations. Figure~\ref{recip} shows measurements of the spin accumulation for different electric and magnetic configurations. We proceed as follows: with the external field applied in the positive $\mathbf{y}$ direction, we set the contacts in the antiparallel configuration. We then measure the precession signal (curve 1). Next, we exchange the current and voltage probes  and we repeat the measurement (curve 2), this time applying the external magnetic field in the $\mathbf{-z}$ direction. Next, with the leads interchanged, we apply the the field in the $\mathbf{-y}$ direction and set the device's magnetic configuration to antiparallel. We then measure the precession signal, again with the external field in the $\mathbf{-z}$ direction (curve 3). We see that curves (1) and (3) are identical. Curve (2) is shifted in magnetic field and its maximum is 2~m$\Omega$ higher than the other two curves. In fact, in the presence of a $\bm{\omega}_{st} \neq \mathbf{0}$, the last term in eq.~(\ref{eq:prec_2D}) is not invariant if we simply exchange $\mathbf{s} \leftrightarrow \mathbf{d}$ without flipping their directions, $\mathbf{s} \rightarrow \mathbf{-s}$ and $\mathbf{d} \rightarrow \mathbf{-d}$. A small \textit{in-plane} stray field of $\approx 20$~mT is enough to account for a difference of 2~m$\Omega$.

We also note that the spin valve traces for all magnetic configurations at zero magnetic field are offset from one another by about 1--2~m$\Omega$. We believe that again the last term of eq.~(\ref{eq:prec_2D}) is responsible, as reversing the electrodes' magnetisations $\mathbf{m}_j \rightarrow -\mathbf{m}_j$, also causes $\bm{\omega}_{st}$ to reverses. We can exclude a Hall effect generated by the leads' magnetic field. In fact, taking the Hall resistance $R_H = -3.5\cdot10^{-11}\ \Omega \cdot m/T$ for Al in the low field limit\cite{ashcroft}, the Hall contribution would be $R_H B/d \approx 0.1$~m$\Omega$, using the thickness of the island $d = 30$~nm and a field of $100$~mT, which is too small to explain the difference.

\begin{figure}
\includegraphics[width = 0.45\textwidth]{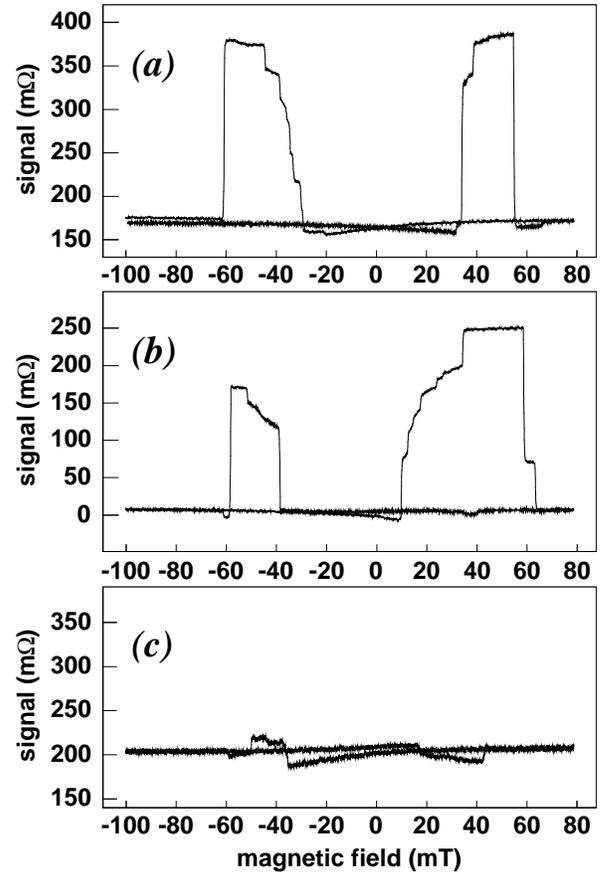}
\caption{Spin valve measurement at 4.2~K in the (a) side, (b) diagonal and (c) opposite configurations. The reversal of the widest electrodes is step-like. In the diagonal configuration, the left and right peak have different height, probably due to incomplete switching of one of the contacts.}
\label{g_meas4K}
\end{figure}

As a comparison, Fig.~(\ref{g_meas4K}) shows measurements at 4.2~K on a different device with tunnel barriers of 2~k$\Omega$ (Al deposition rate 0.2~nm/sec). Side and diagonal configurations show a similar spin signal, around 230~m$\Omega$. In the diagonal case, the left peak does not reach full height, probably due to the incomplete reversal of one of the wide electrodes. The switching of the magnetisation occurs with discrete changes, resulting in a step-like spin signal, as opposed to the switching at RT, which occurs abruptly in most cases. The opposite configuration shows little spin signal, as expected.

We have also measured two devices with tunnel barriers in the range 2--4~k$\Omega$, and found a spin signal of 80~m$\Omega$. For more transparent interfaces, 0.8--1.6~k$\Omega$ (two devices), the spin signal is 55~m$\Omega$. For the last four devices, the spin relaxation time is $\tau_{sf} = 60 \pm 4$~ps. Two devices (Al deposition rate 0.2~nm/sec) were measured both at 4.2~K and at RT: one device, with tunnel barriers of 5--11~k$\Omega$, gave a spin signal of 90~m$\Omega$ at RT and 250~m$\Omega$ at 4.2~K, the other with tunnel barriers 15--35~k$\Omega$, 150~m$\Omega$ at RT and 300~m$\Omega$ at 4.2~K.

We note that the polarisation of the interface decreases from 10.5~\% for the highest resistance interfaces to 8~\% (if we assume that the relaxation time is the same for all devices). Although these values are unequivocally lower than spin polarisation measurements with superconducting aluminium\cite{tedrow}, we see that the aluminium oxide interfaces can be made transparent enough without losing the polarisation completely.

Jedema \textit{et al.}\cite{jedema_02} has performed for Al/Al$_2$O$_3$/Co 1-D structures, spin valve and precession measurements. E-beam evaporation was used (the same evaporating machine) to deposit the metals. They also reported spin flip times of 50~ps but a polarisation of the tunnel barriers at RT of 11~\%, slightly larger that what we found, based on the fit to the experimental traces with a time-of-flight 1--D model that takes the diffusion constant as an independent parameter. They also found, from precession measurements at 4.2~K, that the spin polarisation of the tunnel barriers increases to 13~\% and the spin relaxation time doubles.

As we were not able to perform precession measurements at 4.2~K (due to technical difficulties), we could not determine separately the value of the spin relaxation time and the spin polarisation of the injector and detector. For this reason, we cannot conclude which is the main mechanism of relaxation, whether phonon or impurity induced, and to compare with the theoretical calculations of Fabian and Das Sarma\cite{fabian}. Nevertheless, our spin valve risults at 4.2~K are consistents with those of Jedema \textit{et al.}

\section{CONCLUSION}

Spin accumulation is analysed for zero-dimensional systems, in which the electron spin diffusion time $\tau_{dif\!f}$ is shorter than the spin relaxation time $\tau_{rel}$ and the spin accumulation can be considered uniform. In the system under study, spins are injected into a small island of normal metal through ferromagnetic contacts, and the resulting magnetisation is electrically detected by means of other FM contacts. We have theoretically modelled the island using the finite element theory of Brataas \textit{et al.}\cite{brataas}: we have shown that the presence of the leads affect the spin accumulation by making available extra channels of spin relaxation. In particular, the mixing term $G^{\uparrow \downarrow}$ is selectively relaxing spins with orientation perpendicular to the electrode magnetisation. The expression we derived for the spin accumulation in the island is valid in the case of negligible spin accumulation in the FM contacts.

Experimentally, we have fabricated a small island of aluminium with all dimensions (400~nm$\times$400~nm$\times$30~nm) smaller than the spin relaxation length ($\lambda_{sf} = 550$~nm at RT). Transparent tunnel barriers between the island and the FM electrodes provide a spin dependent resistance that is much higher than all the other (spin independent) resistances in the system, so as to overcome the conductivity mismatch. Because of the lateral dimensions of the island compared to the spin flip length, only pure spin accumulation occurs in our device: the spin signal can therefore be described in terms of the relative orientations of the magnetic electrodes. Spin valve and spin precession measurements were presented for different electrical configurations. The peculiarity in the experiments is that the Ohmic drop across the island is smaller than the spin signal.

In spin valve and precession measurements, we extract the polarisation of the tunnel barriers and the spin flip time. The Al$_2$O$_3$ tunnel barriers, with resistances of $20-100~\Omega\mu \mathrm{m}^2$ still present a certain degree of polarisation, $P = 8\%$. The spin relaxation time at room temperature was found to be $\tau_{sf} = 60$~ps.

The presence of tunnel barriers confine the electrons inside the island and they tunnel out of the system long after having lost their spin information: in fact the escape time is three orders of magnitude larger than the spin flip time. With our lateral devices, it is not possible to directly measure the spin mixing conductance.

\begin{acknowledgments}
The authors want to thank Arne Brataas and Andrei Filip for stimulating discussions, Caspar van der Wal and Julie Grollier for reading the manuscript and Gert ten Brink and Pim van den Dool for technical support. This work was financially supported by MSC$^\textrm{plus}$ and NEDO (Project ``Nano-scale control of magnetoelectronics for device applications'').
\end{acknowledgments}


\begin{thebibliography}{99}
\bibitem{sem} For clarity, we have removed from the micrograph the shadow replicas, as these are not in electrical contact to the device and do not affect the system.
\bibitem{FM_acc} The spin accumulation is assumed to have component along the magnetisation axis, $\bm{\mu}^F = \bm{m} |\bm{\mu}^F|$, because the perpendicular component is rapidly averaged out by the large exchange energy in the ferromagnet.
\bibitem{TBresistances} Note that the tunnel barriers of the narrowest electrodes have higher resistance than the other two: this is due to a small misalignment during metal deposition. The resistances of largest electrodes are equal and so are those of the narrowest ones, though.
\bibitem{anomalous_equivalent} In the equivalent configuration (in the sense of the reciprocity theorem, with current and voltage probes interchanged, and magnetisations reversed), a spin accumulation is present, but it is not detected because the voltage electrodes are parallel to each other.
\bibitem{error} The interpretation we gave in Ref.~\onlinecite{zaffalon} of the difference between antiparallel and anomalous signals as caused by the spin current, is not correct.
\end{thebibliography}
\end{document}